\documentclass[12pt,a4paper]{article}

\newcommand{\ket}[1]{|{#1}\rangle}
\newcommand{\braket}[2]{\langle{#1}|{#2}\rangle}

\newcommand{\dket}[1]{|{#1}\rangle\rangle}
\newcommand{\dbra}[1]{\langle\langle{#1}|}
\newcommand{\osprod}{\prod_{n=1}^{\infty}}

\newcommand{\eq}[1]{Eq.~(\ref{#1})}

\def\beq{\begin{equation}}
\def\eeq{\end{equation}}
\def\beqa{\begin{eqnarray}}
\def\eeqa{\end{eqnarray}}

\newcommand{\EQ}{\begin{equation}}
\newcommand{\EN}{\end{equation}}
\newcommand{\bea}{\begin{eqnarray}}
\newcommand{\ena}{\end{eqnarray}}

\newcommand{\NP}[1]{Nucl.\ Phys.\ {\bf #1}}
\newcommand{\PL}[1]{Phys.\ Lett.\ {\bf #1}}

\newcommand{\PR}[1]{Phys.\ Rev.\ {\bf #1}}

\newcommand{\jhep}[3]{{\it J.~High Energy\ Phys.\ }{\bf {#1}}{\ ({#2})}\ {#3} }
\renewcommand{\thefootnote}{\fnsymbol{footnote}}
\newcommand{\hep}[1]{{\tt hep-th/{#1}}}
\def\one{{\hbox{ 1\kern-.8mm l}}}

\newlength{\bredde}
\def\slash#1{\settowidth{\bredde}{$#1$}\ifmmode\,\raisebox{.15ex}{/}
\hspace*{-\bredde} #1\else$\,\raisebox{.15ex}{/}\hspace*{-\bredde} #1$\fi}
\newsavebox{\uuunit}
\sbox{\uuunit}
    {\setlength{\unitlength}{0.825em}
     \begin{picture}(0.6,0.7)
        \thinlines
        \put(0,0){\line(1,0){0.5}}
        \put(0.15,0){\line(0,1){0.7}}
        \put(0.35,0){\line(0,1){0.8}}
       \multiput(0.3,0.8)(-0.04,-0.02){12}{\rule{0.5pt}{0.5pt}}
     \end {picture}}
\newcommand {\unity}{\mathord{\!\usebox{\uuunit}}}
\newcommand{\Dbar} {{\mbox{\rm$\mbox{I}\!\mbox{D}$}}}

\newcommand {\Cbar}
    {\mathord{\setlength{\unitlength}{1em}
     \begin{picture}(0.6,0.7)(-0.1,0)
        \put(-0.1,0){\rm C}
        \thicklines
        \put(0.2,0.05){\line(0,1){0.55}}
     \end {picture}}}
\newsavebox{\zzzbar}
\sbox{\zzzbar}
  {\setlength{\unitlength}{0.9em}
  \begin{picture}(0.6,0.7)
  \thinlines
  \put(0,0){\line(1,0){0.6}}
  \put(0,0.75){\line(1,0){0.575}}
  \multiput(0,0)(0.0125,0.025){30}{\rule{0.3pt}{0.3pt}}
  \multiput(0.2,0)(0.0125,0.025){30}{\rule{0.3pt}{0.3pt}}
  \put(0,0.75){\line(0,-1){0.15}}
  \put(0.015,0.75){\line(0,-1){0.1}}
  \put(0.03,0.75){\line(0,-1){0.075}}
  \put(0.045,0.75){\line(0,-1){0.05}}
  \put(0.05,0.75){\line(0,-1){0.025}}
  \put(0.6,0){\line(0,1){0.15}}
  \put(0.585,0){\line(0,1){0.1}}
  \put(0.57,0){\line(0,1){0.075}}
  \put(0.555,0){\line(0,1){0.05}}
  \put(0.55,0){\line(0,1){0.025}}
  \end{picture}}
\newcommand{\Zbar}{\mathord{\!{\usebox{\zzzbar}}}}

\begin{document}

%\begin{titlepage}
\begin{flushright} 
hep-th/0002129 \\
KUL-TF-2000/05 \\ 
February 2000
\end{flushright}
\vskip 0.6cm
\begin{center} 
{\Large \bf Boundary states and non-abelian orbifolds\footnote{%
Work supported by the European Commission TMR programme ERBFMRX-CT96-0045.}}
\vskip 0.5cm  
Frederik Roose$^{*,}$\footnote{%
E-mail: {\tt Frederik.Roose@fys.kuleuven.ac.be}} \\
\vskip .2cm
{\sl $^*$ Instituut voor Theoretische Fysica}\\
{\sl Katholieke Universiteit Leuven, B-3001 Leuven, Belgium}
\end{center}
\vskip 0.2cm
\begin{abstract}
In this note the open string orbifold partition function
is analyzed carefully in a way to reveal the group-theoretical aspects. For the simple cases
of ADE orbifolds with regular Chan-Paton action a prescription for consistent boundary states
is given. In the process of outlining how they encode McKay correspondence, we argue that
this results from a non-trivial conspiracy of numerical factors in the string amplitudes.
\end{abstract}

%%%%%%%%%%%%%%%%%%%%%%%%%%%%%%
\renewcommand{\thefootnote}{\arabic{footnote}}
\setcounter{footnote}{0}
%%%%%%%%%%%%%%%%%%%%%%%%%%%%%%%

%%%%%%%%%%%%%%%%%%%%%%%%%%%%%%%%%%%%%%%%%%%%%%%%%%%%%%%%%%%%%%%%%%%%%%%%%%%%%%%%%%
\section{Introduction and summary}
Boundary states encode information about the D-brane open string spectrum and open-closed 
string interactions. Being formulated in a closed string language, they are expected to be
most powerful when trying to extend operations on a closed string theory to the open string
sector in a consistent way. Orbifold theories nicely illustrate this fact, e.g.
the orbifold of type II by $(-)^{F_s}$ to type 0, where in the closed string sector 
space-time fermions are projected out and only diagonally GSO-projected NS-NS and R-R sectors
are kept. After orbifolding the type II boundary states, the number of consistent boundary
states is doubled as prior to the orbifolding. The corresponding type 
0 open string spectrum as predicted in Ref. \cite{KT}
was derived by this method \cite{BCR} . Taking this as an example of a generic
feature one may wish to dispose of boundary states for the more familiar geometric orbifolds.
This was one reason that lead to the present work.

Another piece of motivation comes from geometry.
D-branes in orbifold string theories were first discussed in Ref. \cite{DM}, where among other
issues the spectrum of open string states and open string interactions were derived. The
beautiful interplay between representation theory and the geometry of orbifold singularities is
displayed in the so called McKay correspondence. In the cited reference, the identity
\begin{equation}\label{mckay}
R_Q \otimes R_I = \oplus_{J}\ C_{IJ}\  R_J\ ;
\end{equation}
was taken to basically encode this correspondence, at least for 
two-com\-plex-di\-men\-si\-o\-nal
orbifolds of ADE type. It 
decomposes for you the tensor product of the two-dimensional defining representation and 
an arbitrary irrep of the orbifold group into irreducible representations.  
The multiplicities $C_{IJ}$ 
are encoded in the adjacency matrix of the corresponding extended Dynkin diagram.
Briefly, \eq{mckay} was used in the open string picture : the combined orbifold action on 
the Chan-Paton factor and on the coordinates were used to find the open string massless states,
and hence, the group invariant D-brane configurations. By studying the open-closed and
open-open string interactions, in particular the scalar potential and the D-terms,
Ref. \cite{JM} pointed out in detail that these correspond to
higher-dimensional branes wrapping blow-up cycles. The resolution of the singularity
essentially boils down to the particular hyperk\" ahler quotient construction of Kronheimer
\cite{K}.
These were subject of further study in 
Ref. \cite{DDG}
where they were called 'fractional D-branes'. A physicist's way of thinking about McKay
correspondence may be through this correspondence between blow-up cycles (geometry) and
invariant D-brane configurations (representation theory). 

A natural question that comes up is : how much do closed strings, i.c. boundary states, know
about this surprising correspondence ? Below we try to point out how consistent boundary 
states can be guessed largely from representation theory, with some additional input from
stringy considerations. To point out a peculiar conspiracy between string theory and
representation theory, we put some emphasis first on the general structure of open string
orbifold amplitudes in Section 2. In the next section we make the actual argument. Along the
way, an explicit correspondence between a closed string trace and an open string index
as counting fermions is established.  The original idea was first raised in Ref. \cite{DF},
the heuristic argument going as follows.
Fractional D-branes wrapping the blow-up cycles intersect along so-called 
I-branes \cite{GHM}. The intersection number of cycles is a topological invariant, i.e. invariant
under continuous deformations. Deformed cycles may have additional {\em pairs} of 
intersection points, but these have no net contribution to the intersection number. 
This comes
fairly close to the physical picture of counting chiral (hence massless) fermions. Adding a
chiral-anti-chiral pair does not alter the net chiral fermion number. 

As a by-product of illustrative intentions, we find a prescription for consistent non-abelian
orbifold boundary states. For the $\Dbar_N$ series, this is argued in detail with the result
that a class of consistent boundary states takes the schematic form (see the body text for more
details)
\begin{equation}
\dket{I} = \frac{{\cal N}_0}{\sqrt{|\Gamma|}} \left( \ket{e} + \sum_{g\neq e} |[g]|\ c_g
\chi^I(g) \ket{g} \right)\ ; 
\end{equation}
with coefficients encoding the geometry and representation theory : 
the $\chi^I$ are characters of $\Gamma$, whereas $-c_g^2$ are eigenvalues of the extended
Cartan matrix, i.e. the intersection matrix of blow-up cycles. This in turn provides a new
interpretation of these coefficients. Hitherto, the $c_g^2$ have been interpreted in terms of
Lefshetz numbers \cite{DG}, counting the number of $g$ fixed points that would be obtained 
after toroidal compactification. Our interpretation is {\em local}, as it refers only to the
local resolution of the singularity. It would be interesting to see, however, 
how much of this interpretation survives generalization to higher dimensions.

From the arguments below it should follow in an
obvious way that the prescription also works for the three exceptional groups, although 
no explicit checks were performed.

The question as how to generalize to higher dimensional orbifolds is natural, but
unfortunately, the answer is not. However, we hope that parts of the presented results may 
add to the insight of how strings actually resolve classically singular spaces.

\section{The structure of open string orbifold amplitudes}
Consider first the case of the familiar abelian orbifolds, say $\Gamma =\Zbar_N$ with a
geometrical action on $\Cbar^d$. This action is combined with an action on
the Chan-Paton factor, which we may take to be the regular one. It is a well-known fact that
the regular representation ${\cal R}_{reg} = \oplus_{I}d_I\ {\cal R}_I$, that is, it contains
every $d_I$-dimensional irreducible representation $R_I$ exactly $d_I$ times. 
The group characters
implement the action on the Chan-Paton factor most conveniently, and the $\Gamma$-projected
one-loop partition function is
\begin{equation}
{\rm Tr}_{IJ} (\frac 1 {\Gamma} \sum_g \hat g e^{-2 t H_o}) = \frac {d_I\, d_J} {\Gamma} 
\sum_g 
\bar\chi^I(g) \chi^J(g) {\rm tr}_{\alpha}(g e^{-2 t H_o})\ ;
\end{equation}
for unitary representations. Indices $I,J$ label 
the $d_I \times d_J$ blocks in the Chan-Paton matrix  
. The factor $d_I\, d_J$ in the r.h.s. implements the multiplicities 
in the decomposition of the regular representation.
Further, ${\rm tr_\alpha}$ is a trace in the 
Fock space of
open string oscillator states. The $\Gamma$ action in this space is
twofold : a group element rotates the oscillators along the orbifold directions but may also
affect the ground state. We may represent the geometrical action of $g$ on the string fields
by commuting diagonal $SU(d)$ rotations, 
\begin{equation}
{\cal Z}^i \rightarrow \omega^{q_i} {\cal Z}^i\ ,\ i=1\ldots d\ ; \omega = e^{\frac{2\pi
i}{N}}\ ;
\end{equation}
where ${\cal Z}^i = Z^i + \theta \psi^i$ a worldsheet superfield.
Being a scalar the NS-sector ground state remains inert under the rotations, while the
action of $g$ on the R-spinor ground state is
\begin{equation}\label{spinrot}
\left[\prod_{i=1}^d \left( \cos \frac{\pi q_i}{N}
\right) \ \unity + ({\rm traceless})\right] \ket{0}_{\rm R}
\ ;
\end{equation}
In the Fock space trace a factor $16$ combines with the products of cosines. This numerical
value is the number of non-GSO projected on-shell spinor states in the Ramond sector,
multiplied by a weight from the spinor rotation. This number was interpreted in Ref. \cite{DG}
as implementing the true action of $g$, where this should only be understood as an "effective"
action, i.e. inside a spinor trace.

The structure in \eq{spinrot} is motivated by decomposing the spinor in 
tensor products of two-dimensional spinors. E.g. 
when a single complex coordinate gets multiplied by a phase
$e^{2 \pi i q/N}$, the corresponding two-dimensional spinor is acted upon by 
$\exp (\pi i q/N \sigma_3)$ yielding the stated result.

As will become clearer soon, it may be instructive to consider the detailed structure of the 
Fock space traces in some detail. For the remainder of this paper we
have chosen to adopt
the conventions of Ref. \cite{DG}. 

For convenience, let us focus on the ${\rm tr}_{NS}$ part, 
temporarily suppressing
the ${\rm tr}_{NS}(-)^F$ and ${\rm tr}_R$ contributions. 
The trace over the oscillators is easily seen to yield
\begin{equation}
q^{-1/2}
\left[\frac{\osprod (1 + q^{n-1/2})}{\osprod (1 - q^n)}\right]^{8-2d}
\prod_{i=1}^d\left[
\frac{\osprod (1 + z_i q^{n-1/2})(1 + \bar z_i q^{n-1/2})}{\osprod (1 - z_i q^n)(1- \bar z_i 
q^n)}
\right]\ ;
\end{equation}
where $z_i = e^{2 \pi i\nu_i}$, with $\nu_i = q_i$, and $q = e^{-2 \pi t}$.
Equivalently, this is rewritten in terms of standard $\theta$-functions
\begin{equation}
2^d \prod_{i=1}^d \sin\pi\nu_i \left[\frac{\theta_3(0|it)}{\eta(it)^3}\right]^{4-d}
\prod_{i=1}^d \frac{\theta_3(\nu_i|it)}{\theta_1(\nu_i | it)}\ . 
\end{equation}
Similarly, for the other spin structures one finds
\begin{eqnarray}
{\rm tr_{NS}}(g (-)^F e^{- 2 t H_o}) &=& 
-2^d \prod_{i=1}^d \sin\pi\nu_i \left[\frac{\theta_4(0|it)}{\eta(it)^3}\right]^{4-d}
\prod_{i=1}^d \frac{\theta_4(\nu_i|it)}{\theta_1(\nu_i | it)}\ ; 
\\ \label{theta2}
{\rm tr_{R}}(g e^{- 2 t H_o}) &=&
2^d \prod_{i=1}^d \sin\pi\nu_i \left[\frac{\theta_2(0|it)}{\eta(it)^3}\right]^{4-d}
\prod_{i=1}^d \frac{\theta_2(\nu_i|it)}{\theta_1(\nu_i | it)}\ . 
\end{eqnarray}
Notice that the $\cos \pi\nu_i$ factors in the expansion of $\theta_2(\nu_i|it)$ correctly take
into account the rotation of the spinor ground state by $g$. The factor $2^d$ and the product
of sines are added such as to cancel $\theta_1$ factors that are spurious in the open string
picture. Furthermore, the Ramond sector ground state degeneracy $2^4$ and the rotation cosine
factors are correctly accounted for in the $\theta_2$.

For future reference, let us look at the open string massless level. Extracting the massless
$q^0$ piece,
we find
\begin{eqnarray} 
{\rm Tr}_{IJ,NS}(\frac 1 {|\Gamma|}\sum_g \hat g\frac{1+(-)^F} 2) &=&
\frac {d_I\, d_J} {|\Gamma|}\Big[ 8 + \sum_{g\neq e} \bar\chi^I(g)\chi^J(g) [(8-2d) + \sum_{i=1}^d(z_i+\bar
z_i)]\Big]\ ; \nonumber \\
\label{tropenns} && \\ \label{tropenr}
{\rm Tr}_{IJ,R}(\frac 1{|\Gamma|}\sum_g \hat g\frac{1+(-)^F} 2) &=&
\frac {16\ d_I\, d_J} {2 |\Gamma|} \sum_g \bar\chi^I(g)\chi^J(g)[2^d\prod_{i=1}^d \cos
\pi\nu_i]\ ; 
\end{eqnarray}
In the latter equation, the factor 16 is due to the spinor trace and the 2 in the 
numerator is due to the GSO projection.
The structures of these equations reflect the different transformation properties of scalars 
and vectors resp. spinors under spacetime rotations. A detailed account of the abelian orbifold
can be found in Ref. \cite{DG}. We have chosen to display only those features which are
relevant for the study of the non-abelian case, to which we turn next.

A simple example of a non-abelian orbifold is provided by taking $d=2, \Gamma=\Dbar_N$. 
This discrete subgroup of $SU(2)$ is generated by 2 rotations $a,b$ of orders $2N$ and $4$ 
respectively. They satisfy the further relations $a^N\,b^2 = e$ and $bab^{-1}=a^{-1}$.
The conjugacy classes are
\begin{equation}
\{e \}\ ;\ \{a^N\}\ ;\ \{a^k,a^{-k}\}\ ;\ \{ba^{2m}\}\ ;\ \{ba^{2m+1}\}\ ;
\end{equation}
where $k=1\ldots N-1$ and $l=0\ldots N-1$. Further, there are $N-1$ two-dimensional irreps, 
in one-to-one correspondence with the two-element conjugacy classes, and four one-dimensional
irreducible representations. Finally, 
the defining representation is given by the $SU(2)$ matrices
\begin{equation}\label{drot}
A = \left( \begin{array}{cc}
\omega & 0 \\
0 & \omega^{-1}
\end{array} \right) \ ;
\quad
B = \left( \begin{array}{cc}
0 & i \\
i & 0
\end{array} \right) \ ; \quad \omega = e^{\frac{\pi i}{N}}\ ;
\end{equation}
Define now ${\cal Z}^i =
Z^i+\theta \psi^i$ and ${\cal W}^i = W^i + \theta\lambda^i,i=1,2$, as worldsheet complex 
superfields along the orbifold directions. The ${\cal Z}^i$ are coordinates such that the
rotations $A^k$ are diagonal. Likewise, in the ${\cal W}^i$ coordinate basis, it are the
$BA^k$ that are diagonalized to $i\sigma_3$, 
the latter being $\Zbar_4$ rotations. The one-loop open string amplitude is then decomposed
accordingly. For brevity only displaying the R part before GSO-projection, we find
\begin{eqnarray}
{\rm Tr_{IJ,R}}  &=& \frac{d_I\, d_J}{4\ N}\Big[ \sum_{k=0}^{2N-1} \bar \chi^I(a^k)\chi^J(a^k) 
{\rm tr_{R,{\cal Z}}}(a^k) 
 \nonumber \\
&+& \sum_{k=0}^{N-1} \bar \chi^I(ba^{2k})\chi^J(ba^{2k}) {\rm tr_{R,{\cal W}}}(ba^{2k})
\nonumber \\ \label{ortrace}
&+& \sum_{k=0}^{N-1} \bar \chi^I(ba^{2k+1})\ \chi^J(ba^{2k+1}) {\rm tr_{R,{\cal W}}}(ba^{2k+1})
\Big] \ .
\end{eqnarray}
With the characters depending only on the conjugacy class $[g]$ of the element $g$. 
\eq{ortrace} and \eq{theta2} yield\footnote{Actually, the second term is zero, as
$\theta_2 (1/2|it)$ vanishes.}
\begin{eqnarray}
&& \frac{d_I\, d_J}{4\ N} \left(\frac{ \theta_2(0|it)}{\eta^3(it)}\right)^2
\Big[ \left(\frac{ \theta_2(0|it)}{\eta^3(it)}\right)^2
+ 4\bar\chi^I(a^N)\chi^J(a^N) \sin^2\frac \pi 2 
\left(\frac{\theta_2(\frac 1 2|it)}{\theta_1(\frac 1 2 | it)}\right)^2
\nonumber \\
&& + \sum_{k=1}^{N-1} 4 |[a^k]|\bar \chi^I(a^k)\chi^J(a^k)\ \sin^2 \frac {\pi k} {2N} 
\left(\frac{\theta_2(\frac{k}{2N}|it)}{\theta_1(\frac{k}{2N} | it)}\right)^2
\nonumber \\
&& + 4 |[b]| \bar \chi^I(b)\chi^j(b)\ \sin^2 \frac \pi 4
\left(\frac{\theta_2(\frac 1 4|it)}{\theta_1(\frac 1 4 | it)}\right)^2
\nonumber \\ \label{dntropen}
&& +  4 |[ba]| \bar \chi^I(ba)\chi^j(ba)\ \sin^2 \frac \pi 4
\left(\frac{\theta_2(\frac 1 4|it)}{\theta_1(\frac 1 4 | it)}\right)^2
\Big] \ .
\end{eqnarray}
Extract the massless part,
\begin{eqnarray}
\frac 1 {4N} \Big[\ 8 &+& \sum_{k=1}^{N-1} 
4.2 \ \bar \chi^I(a^k)\chi^J(a^k)\ \cos^2 \frac {\pi k} {2N} 
\nonumber \\
&+& 4.N \ \bar \chi^I(b)\chi^J(b)\ \cos^2 \frac {\pi} {4} 
\nonumber \\
&+& 4.N \ \bar \chi^I(ba)\chi^J(ba)\ \cos^2 \frac {\pi} {4} \Big]\ ;
\end{eqnarray}
Gathering the massless parts of ${\rm Tr}_{IJ,R}$ in a matrix $D_{IJ}$, it is found that
its $(I,J)$ entries are the {\em absolute values } of the corresponding entries in
the extended Cartan  matrix $\hat D_{N+2}$.
This was to be expected from the discussion 
of Ref. \cite{DM} from which it is seen to be a mere consequence of
group representations and supersymmetry. Also,
this gives a detailed account of the connection between the Ramond sector traces $D_{IJ}$
and an
intersection matrix $\hat D_{N+2}$ as an index, as alluded to in Ref. \cite{DF}. 

\section{Boundary states}
It is now rather obvious how to construct the boundary states that reproduce this open string
spectrum. Write the desired boundary states schematically
\begin{eqnarray}\label{globbs}
\dket I \hspace{.8cm} &=& \frac {{\cal N}_{o}}{\sqrt{2 |\Gamma|} }\Big[(\ket{+}_{NS} - \ket{-}_{NS}) + 
4i\ (\ket{+}_R + \ket{-}_R) \Big] \ ; \\ 
\label{detbs}
\ket{\pm}_{NS/R} &=& \ket{e,\pm} + {\cal N}_{NS/R} \sum_{g\neq e}  c_g\ \chi^I(g)\
 \ket{g,\pm}_{NS,R}\ ; 
\end{eqnarray}
GSO invariance imposes consistency constraints. Considering only BPS Dbranes, we borrow a
result from the analysis in Ref. \cite{BG}. Thus the sign of ${\cal N}_{NS/R}$ in \eq{detbs} is
fixed to be $+1$.  
The overall normalization ${\cal N}_o$ is universal, and is fixed by comparison of 
${\rm Tr_{NS}}$ in
the open string and the modular transformed sum of untwisted $_{NS}\braket{e,\pm}{e,\pm}_{NS}$
contributions. 
This program was carried out in Ref. \cite{DG}, and can be taken over here,
and we do not repeat the calculation here. The only difference is that
this overall normalization constant found there has to be multiplied by $\sqrt{d_I\, d_J}$, as
follows from \eq{tropenr}.

To completely specify $\dket{I}$ we need only fix the coefficients $\{c_g\}$. Of course, one
way to proceed is to do compute the closed string tree level exchange in each twist sector and
make it match with the corresponding projection sector in the open string one loop amplitude.
Ultimately, this yields consistent boundary states by construction. 
Instead however, the presentation below will turn this logic around :
we first try and guess the right answer, and only a posteriori do we verify the claim by
the above procedure. 
To make an educated guess, first observe that in the abelian case of $\Gamma = \Zbar_N$ with
one generator $g$, say, these coefficients were found to be \cite{BCR} $c_{k} = 2 \sin 
\frac{ \pi k} {N} $, where $k$ denotes the power of the generator. 
In what sense could these be interpreted such as to allow for easy
generalization ? The basic point here is that the columns of the character table $\chi^I(k)$
are eigenvectors of the extended Cartan matrix of $(\hat A_{N-1})^I_J$. 
The corresponding eigenvalues are $c_{k}^2$, where $k=1\ldots N-1$ and zero for 
$\chi^I(e)$. Now this 
is the formulation that is suitable for generalization to the non-abelian case. 

For the $\Dbar_N$ orbifold, pick the extended Cartan matrix of $D_{N+2}$ :
\begin{equation}
{\left(\hat D_{N+2}\right)^I}_J = \left(
\begin{array}{ccccccccc}
-2 & 0 & 1 & 0 & \cdots &&& & 0 \\
 0 & -2 & 1 & 0 &&&&&  \\ 
 1 & 1 & -2  & 1 & & &&&  \\
 0 & 0 & 1  & -2 & 1 & 0 &&& \vdots  \\
 \vdots & & & 1  & -2 & 1 & && \\
 & & & & 1 &  \ddots & & &  \\
&&&&&&1&& \\
  &&&&& 1 & -2 & 1 & 1 \\ 
  &&&&& 0 & 1 & -2 & 0 \\
 0 & &&& \cdots & 0 & 1 & 0 & -2 
\end{array} \right) \ ; 
\end{equation}
One may check
\footnote{ The easiest way to do the job, is by tracing \eq{mckay}, thus immediately providing
the spectrum of the connectivity matrix. } 
that the character table of $\Dbar_N$ still enjoys the property that its columns
are eigenvectors of ${\left(\hat D_{N+2}\right)^I}_J$. Moreover, labelling them by the conjugacy
classes $\big[g\,\big]$, the spectrum of corresponding eigenvalues is found to be 
$\{-4 \sin^2 \pi q_{[g]} \}$, where $q_{[a^k]} = \frac{ k}{2 N}$, with $ k=0 \ldots N$, while
$q_{[b]} =  q_{[ba]} = \frac{1}{4}$. The zero eigenvalue corresponds to the vector of
characters evaluated on the neutral element. In view of what follows, 
we denote the eigenvalues by $-c^2_{[g]}$. They
can be arranged in a diagonal matrix, with non-zero entries $|[g]| c^2_{[g]}$ such that
\begin{equation}\label{grorth}
-\frac 1 {|\Dbar_N|}\sum_{[g]} |[g\, ]|\  c_{[g]}^2\  \bar\chi^I([g])\ \chi^J([g]) =
{\left(\hat D_{N-2} \right)^I}_J\ .
\end{equation}
Observe 
that the values of the characters evaluated on $[e]$ drop out of this equation. 
As an aside, it may be noticed that adding
\begin{equation}\label{add1}
\frac 4 {|\Dbar_N|} \sum_{[g]} |[g]|\ \bar\chi^I([g])\ \chi^J(g) = 4 \ {\delta^I}_J
\end{equation}
to both sides of \eq{grorth} turns the sines on the left-hand side into cosines, or equivalently, it
shifts the $-2$'s on the diagonal of the Cartan matrix into $+2$ values. 

What do we buy from this juggling with numbers ? Inspired by the formal similarity between
\eq{dntropen} and \eq{grorth}, the guess is put forward that the
coefficients $c_{g} = c_{[g]}$ in front of the Ishibashi components of \eq{detbs} defines
consistent boundary
states; that is, they can be verified to give rise to proper open string channel amplitudes
upon modular transformation. As such, they satisfy Cardy's condition\ \cite{cardy}. 

Let us point out some salient features of this construction. 
First, the conjugacy classes  are known to be in one-to-one
correspondence with orbits of closed string twisted sectors under the orbifold projection. As
such, the sums over the conjugacy classes in \eq{grorth} should translate accordingly. As
ultimately the extended Cartan matrix is closely related to the massless spectrum of 
open strings \cite{DM}, Eqs.~(\ref{grorth}) and (\ref{add1}) are to find some open
string equivalent operation. Of course, the sums cannot be but possibly implementing the
orbifold projection. Hence, they run over projection sectors contributing to the partition
function with insertions of $\hat g$. In turn, these open string sectors arise from closed
string corresponding twist sectors. As an example, in \eq{grorth}
$[e]$ can be thought of as labelling both the unprojected open string and the untwisted
closed string sectors.
In particular, from \eq{add1} where the 
non-vanishing contribution of $[e]$ shows up, a remarkable interplay between
the relative normalizations in front of $\ket{0,\pm}$ and $\ket{g \neq e,\pm}$ has to be
suspected ! 

To appreciate this fact, one has to work one's way properly through the closed 
string computation in principle. This is straightforward but tedious. Instead,
in order that technical details do not blur the point we wish to make, we suppress inessential
integrals over the closed string modulus $l$, and global factors. From 
the bare essentials the r\^ ole of the group-theoretical coefficients will be clarified. For
details of the calculation, we refer to Ref. \cite{DG}, treating the analogous threefold
abelian orbifold case. 
  
With a proper choice of coordinates, the geometric
action of any group element can always be diagonalized, as pointed out in the previous section.
Depending on the type of 
twisted sector it is then more appropriate to choose one set of oscillators or the other.
As such the treatment of the twisted sectors differs in no respect from the abelian case,
and the corresponding Ishibashi states follow at once from Ref. \cite{DM}.
We do not display them for brevity.

These Ishibashi components of a state $\dket I$  
yield for the exchange in e.g. the NS-NS $a^k$-twisted sector 
(omitting prefactors in full as promised) 
\begin{eqnarray}
_{NS}\braket{a^k,+}{a^k,+}_{NS} &=& \left(\frac{\theta_3(0|2il)}{\eta^3(2 il)}\right)^2 
\left( \frac{\theta_3(\frac{kl}{2N}| 2il )}{\theta_1(\frac{kl}{2N} | 2il)}\right)^2
\ ; \\
&\rightarrow& \left(\frac{\theta_3(0|it)}{\eta^3(it)}\right)^2 
\left( \frac{\theta_3(\frac{k}{2N}| it )}{\theta_1(\frac{k}{2N} | it )}\right)^2
\ ; 
\end{eqnarray}
Similarly,
\begin{eqnarray}\label{nsnspp}
_{NS}\braket{a^k,-}{a^k,+}_{NS} &=& \left(\frac{\theta_4(0|2il)}{\eta^3(2 il)}\right)^2 
\left( \frac{\theta_4(\frac{kl}{2N}| 2il )}{\theta_1(\frac{kl}{2N} | 2il)}\right)^2
\ ; \\ \label{nsnspm}
&\rightarrow& \left(\frac{\theta_2(0|it)}{\eta^3(it)}\right)^2 
\left( \frac{\theta_2(\frac{k}{2N}| it )}{\theta_1(\frac{k}{2N} | it)}\right)^2\ ; 
\end{eqnarray}
The symbol $\rightarrow$ denotes the modular transformation $l \rightarrow 1/2t$ to the open
string channel.
From the definition of $\theta_{1,2}$
\begin{eqnarray*}
\theta_1(\frac {\pi k} {2N}| it) &=& 2 \exp(-\pi t/4) \sin (\frac{\pi k}{2N}) 
\osprod (1-q^n)(1-z q^n)(1-\bar z q^n)\ ; \\
\theta_2(\frac {\pi k}{2N}| it) &=& 2 \exp(-\pi t/4) \cos (\frac{\pi k}{2N}) 
\osprod (1-q^n)(1+z q^n)(1+\bar z q^n)\ ;
\end{eqnarray*}
where $q,z$ are defined as in Section 2.
A peculiarity of $\theta_{1,2}$  as opposed to their 
$\theta_{3,4}$ counterparts, is that they come with factors $ 2\sin \pi \nu$ and 
$2\cos \pi\nu$
respectively, in front of the triple infinite products (from the "oscillators" \footnote{With
abuse of language, because one of the infinite product factors, $\osprod (1 - q^n)$, does not
actually correspond to a physical oscillator. Rather, it cancels with a similar factor in the
numerator.} ). 
 
We are finally in a position to fully display a remarkable interplay of numerical factors.
In, say the closed string exchange between branes $\dbra I$ and $\dket J$,
the contributions from the $_{NS}\braket{g,\pm}{g,\pm}_{NS}$, add up according to \eq{detbs},
yielding at the massless open string level
\begin{equation}
\frac {d_I\, d_J} {|\Dbar_N|} \Big[{\rm tr_{NS}} + {\cal N}_{NS}^2 \sum_{g \neq e} 4\sin^2 \pi q_g\ 
\bar\chi^I(g)\ \chi^J(g) {\rm tr_{NS}}\ g\ \Big]\  ;
\end{equation}
From Section 2, it is clear that the Fock space traces yield combinations of
$\theta_3,\theta_1$ and $\eta$, upon expansion of which one finds at the massless level
\begin{equation}
\frac {d_I\, d_J} {|\Dbar_N|} \Big[\ 8 + {\cal N}_{NS}^2
\sum_{g\neq e} \bar\chi^I(g)\chi^J(g) [\ (8-2d) + \sum_{i=1}^d(z_i+\bar z_i)]
\Big] \ ;
\end{equation}
Tuning ${\cal N}_{NS}^2$ to 1 is necessary to obtain a consistent open string interpretation, as
in \eq{tropenns}. 
To make it work, $4 \sin^2$  factors from the boundary state 
precisely cancel those from the $\theta_1$ factors in the denominator of the twisted sector
Fock space traces.

As to the Ramond sector, the situation is similar.
Being entirely due to the $_{NS}\braket{g,\pm}{g,\mp}_{NS}$ summands in $\dbra I$ and
$\dket J$, one obtains from \eq{detbs}
\begin{equation}
\frac {d_I\, d_J} {|\Dbar_N|} \Big[{\rm tr_R} + {\cal N}_{NS}^2 \sum_{g \neq e} 4\sin^2 \pi q_g\ 
\bar\chi^I(g)\ \chi^J(g) {\rm tr _R}\ g\ \Big]\  ;
\end{equation}
The value of ${\cal N}_{NS}^2$ has been set to 1 above; upon expansion of the traces,
\eq{tropenr} is exactly reproduced. Again, the necessary cosine and sine factors, 
and a factor cancelling the 4 in the boundary state $c_g$, are provided by the
$\theta$-functions.

It is tempting to interpret the boundary state coefficients in combination with the 
characters as intersection numbers of the cycles resolving the orbifold space. With this input
from geometry, we find it quite a curious fact that they combine appropriately with the
typically stringy $\theta$ functions such as to yield a proper open string partition function.

Turning things around, it is quite remarkable that it are eigenvalues of the 
intersection matrix that enter exactly as coefficients of the Ishibashi states. 
In retrospect, however, these
features should find their origin in the McKay correspondence relating \eq{mckay} with the
geometry.

\bigskip
\subsection*{Acknowledgments}
I wish to thank Ben Craps for enlightening discussions on this and related subjects. Marco
Bill\' o is greatly acknowledged for useful comments on the manuscript.


\begin{thebibliography}{99}
\bibitem{KT}
I.R.~Klebanov and A.A.~ Tseytlin, {\it D-branes and dual gauge theories in type 0 strings}, 
\NP{B546} (1999) 155, \hep{9811035}.

\bibitem{BCR}
M.~Bill\' o, C.~Craps and F.~Roose, {\it On D-branes in type 0 string theory}, 
\PL{B 457} (1999) 61, \hep{9902196}.

\bibitem{DM}
M.R. Douglas and G. Moore, {\it D-branes, quivers and ALE instantons}, 
{\tt hep-th/9603167}.

\bibitem{JM}
C.~Johnson and R.~Myers, {\it Aspects of type IIB theories on ALE spaces}
\PR{D55} (1997) 6382, \hep{9610140}.

\bibitem{K}
P.B.~Kronheimer and H.~Nakajima, {\it Yang-Mills instantons on ALE gravitational instantons},
Math. Ann. {\bf 288} (1990) 263.

\bibitem{DDG}
E.~Diaconescu, M.R.~Douglas and J.~Gomis, {\it Fractional branes and wrapped branes},
\jhep{02}{1998}{013}, \hep{9712230}. 

\bibitem{DF}
M.~Douglas and B.~Fiol, {\it D-branes and discrete torsion II}, \hep{9903031}.

\bibitem{GHM} 
M.~Green, J.~Harvey and G.~Moore, {\it I-brane inflow and anomalous couplings on
D-branes}, Class. Quant. Grav. {\bf 14} (1997) 47, \hep{9605033} .

\bibitem{BG}
O.~Bergman and M.R.~Gaberdiel, {\it On the consistency of orbifolds}, \hep{0001130}.
\bibitem{DG}
E.~Diaconescu and J.~Gomis, {\it Fractional branes and boundary states in orbifold theories},
\hep{9906242}.

\bibitem{cardy}
J.L.~Cardy, {\it Boundary conditions, fusion rules and the Verlinde formula}, \NP{B324} (1989)
581, 


\end{thebibliography}
\end{document}